# Low-cost mobile 3D scanning of heritage objects to facilitate long-distance research collaboration

*Dirk HR Spennemann [1] and Sharnie Hurford[2]*

[1] Gulbali Institute; Charles Sturt University; PO Box 789; Albury NSW 2640, Australia.
Correspondence: dspennemann@csu.edu.au

[2] School of Indigenous Australian Studies; Charles Sturt University; Bathurst NSW 2795; Australia

*

**Abstract:** While three-dimensional visualization has become a common tool in various cultural heritage applications, the emphasis has been on high fidelity representation, essentially the generation of digital twins or digital reconstructions. Overlooked appears to be the utility of 3D in research collaboration where one of the researchers has access to the original, potentially fragile object while the others are based in remote locations. This paper describes the application of a low-cost, mobile and swiftly executable 3D scanning process and discusses the benefits of this for remote collaboration of three-dimensional objects of material culture.

**1. Introduction**

Three-dimensional visualization has become a common tool for the representation of cultural heritage objects in museum settings (Luther, Baloian, Biella, & Sacher, 2023), to create interactive, web based virtual museums (Carvajal, Morita, & Bilmes, 2020; Sillaurren & Aguirrezabal, 2012; Zidianakis et al., 2021), to generate 'digital twins' (Qian et al., 2022) for heritage interpretation (Dezen-Kempter, Mezencio, Miranda, De Sãi, & Dias, 2020) , but also specifically to create visualizations in case of destruction during natural disasters or armed conflict (Hou, Lai, Wu, & Wang, 2024; Shabani et al., 2022; Vuoto, Funari, & Lourenço, 2023, 2024), as well as gaming environments representing historic settings (Champion, 2016; Ishar, Zlatanova, & Roberts, 2022; Kargas, Loumos, & Varoutas, 2019). These 3D models may be static but are more commonly animated. Some of these visualizations represent the objects as they are (Girelli, Tini, D'apuzzo, & Bitelli, 2020), while others represent reconstructions of earlier, pre-modification, pre-decay or pre-disaster states (Martinez Espejo Zaragoza, Caroti, & Piemonte, 2021). Given the desired quality of output, these approaches commonly require photogrammetric equipment (for larger sites) or laboratory-/studio-based set ups with controlled lighting and camera/recording positions.

Despite the plethora of papers discussing actual and possible uses of 3D models in cultural heritage studies, there appears to be no mention of their use in research collaboration. This paper describes a low-cost, mobile 3D scanning process used to represent three-dimensional objects of material culture for collaborative research by researchers separated by large distances and discusses the benefits of 3D models for remote collaboration.

*1.1. Background*

The authors are generally interested in the attitudes of the dominant Australian community of the 1950s and 1960s towards Indigenous Australian peoples and how a then essentially a settler-colonialist mindset manifested itself in the stereotyping of appearances of Indigenous Australian peoples and in the appropriation of Indigenous Australian motifs on objects of post-World War II material culture (Spennemann, 2022; Spennemann & Hurford, subm.-a; Spennemann & Singh, 2023). One of the prominent examples are small slip-cast ceramic pieces ('pin dishes') created by Vandé pottery (Sydney) that exhibit the heads of Indigenous Australian men in relief. In the preparation of an academic manuscript discussing these pin dishes (Spennemann & Hurford, subm.-b), it was essential for both authors to be able to examine, analyze and discuss these pieces and in particular the 'physiognomy' of the relief heads. This was, however, encumbered by the fact that the authors reside on different campuses of the same academic institution spaced 450˚km apart. An appropriate solution was considered to be the generation of 3D models.

*1.2. Requirements*

For the purposes of the research collaboration there was no need to develop high-quality but time-consuming models suitable for external presentation and public consumption as these models were to be solely used for internal discussion. The requirements were that the process

- had to be low cost
- could be executed with personally available technology,
- was easy to learn,
- could be implemented in a non-laboratory setting, and
- was not time consuming to execute

## 2. Technique

The plates were scanned using the rear camera of one of the author's (DHRS) standard personal Samsung S21 Android smartphone and converted into a 3D model with the KiriEngine Application (free version) with server-based processing (KIRI Innovation, 2024b). The process entailed to place the plate on a table in ambient daylight and to take 100 sequential images of the object with the KiriEngine Application (which draws on the camera's 12 megapixel setting). As the focus was on the relief heads of the pottery, the backside of each plate was irrelevant and thus not scanned. The application uploads the images to the vendor's servers, where they are queued for processing. The application window needs to remain open and in the foreground during the upload process, the duration of which is subject to network speed and server responsiveness. Once the upload is completed, the user regains full access to all smart phone functions without impact on the model. Once processing is complete, the 3D model will reside on the KiriEngine server (until user deleted) and can be viewed both within the smartphone application itself or on any web browser using a shared weblink.

While not necessary for the functionality required for the collaborative research, the authors chose to 'tidy up' the scans by removing egregious excess areas (i.e. parts of the table-top included in the scan when moving the camera). The trimming entailed to interactively move a cropping slider to the required position and to save that crop as a new version—and to repeat the process until all four sides had been cropped resulting in the plate residing on a rectangular 'base'. A copy of the final plate model was then further cropped to isolate the head.

None of the image generation processes require operator intervention and therefore can run in the background. Sound feedback alerts the user once the model processing (initial generation or trimmed models) is complete.

## 3. Results

For the project, five plates, representing different molds, were scanned and processed on 4 December 2024 (Appendix A). For this paper, an additional five plates were scanned and processed (7 and 8 December), noting the required time for image acquisition, uploading, processing and final cropping (Appendix B). The recorded times are indicative only as they depend on the efficiency of the operator (image acquisition, adjusting cropping boundaries), network connections (upload speeds) and server load (remote processing). The impact of the server load, for example, is exemplified by the remotely executed processing and saving of a cropped 3D model which took between 41 and 9 minutes 9 seconds (average 5 minutes 13 seconds, n=16). Differential load on the KiriEngine server, which can only process 12 scans at any given time (KIRI Innovation, 2024a), also impacted the generation of the original model which ranged from 11 minutes 52 seconds to 49 minutes (Table 1).

*Table 1: Time investment in the key phases of 3D model generation*

|  | Average | Min–Max |
|---|---|---|
| Image acquisition | 03:11±00:21 | 02:45–03:38 |
| Upload | 12:16±05:12 | 07:52–19:52 |
| Processing | 27:46±17:30 | 13:50–49:00 |
| Cropping to Plate | 19:28±09:34 | 05:21–30:38 |
| Cropping to Head | 18:03±09:34 | 04:34–29:02 |
| File renaming / deletion | 00:13±00:02 | 00:10–00:15 |
| Total | 1h20:46±18:27 | 1h 01:24–1h 47:38 |

The completed models reside on the KiriEngine server and c n be accessed via server-assigned URLs (Appendix A). In addition, the application allows for download of 3D model data for external archiving. The models used for the main study (Spennemann & Hurford, subm.-b) as well as this paper have been archived in the research repository of the authors' institution.

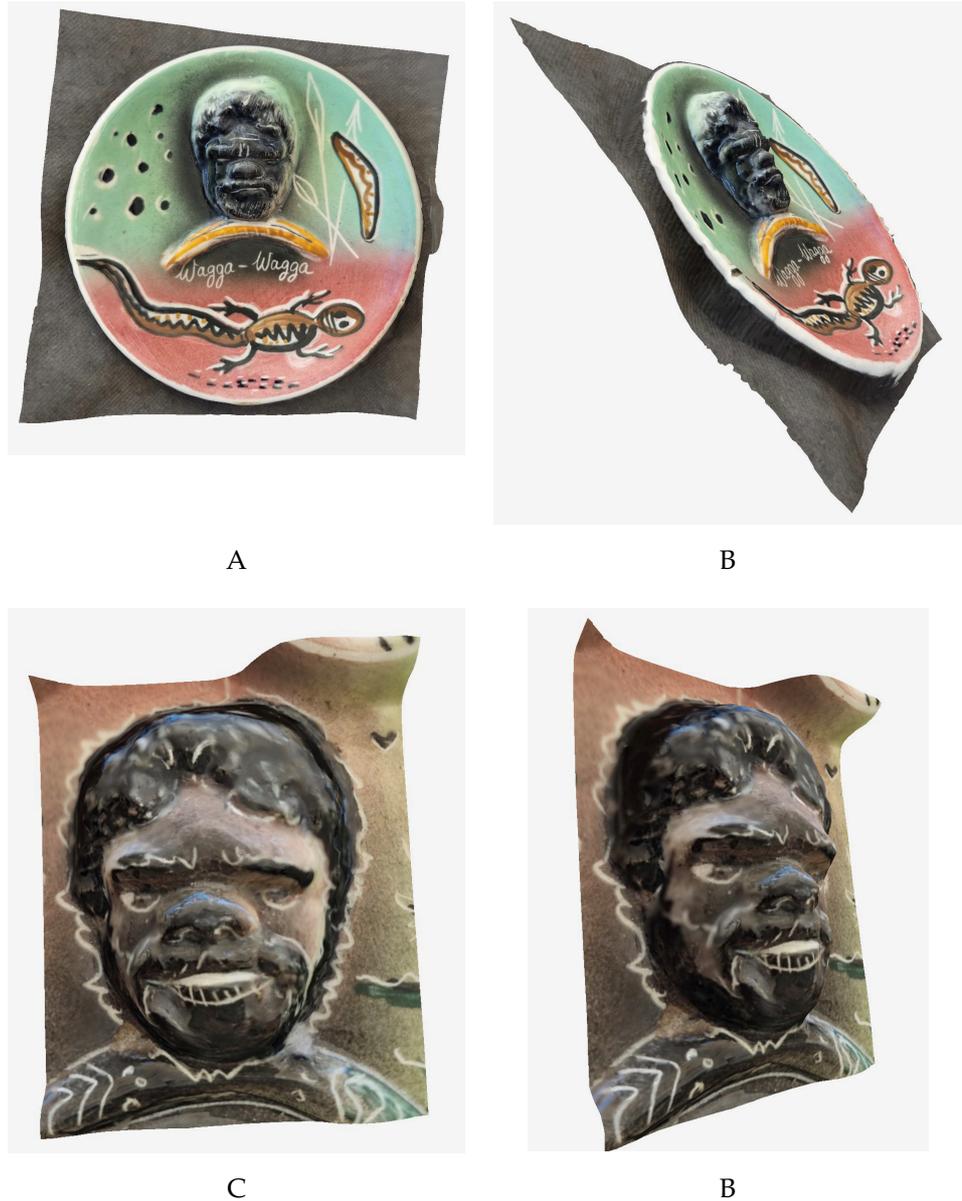

*Figure 1.* Screen shots of two 3D models of Vandé pin dishes generated for the project.
A, B) plate A2-7 (whole plate); C, D) plate B1 (head section only).

## 4. Discussion

As noted, the purpose of the 3D scanning was not to create high-quality models for presentation, but to rapidly and comparatively effortlessly generate 3D models that could be repeatedly viewed by a second, remotely located researcher. The direct benefits of the approach were that the images could be generated using personally available (extant) smartphone technology in a non-laboratory setting and that the generation of the models was not time consuming. Server-based access to the 3D model meant that the second author could manipulate the object virtually, thus obviating inter-campus travel with its costs, effort, travel time and carbon footprint. Additionally, it reduced the need for repeated handling of the original, potentially fragile objects, which also applied to the first author.

An unanticipated benefit of the 3D model was that the images were rendered without the gloss of the ceramic glaze. While this would be considered a 'fault' in 3D models used for formal public representation, it allowed to examine the 'physiognomy' of the heads without the distracting reflections that the glaze generated on the dark colored faces of the originals.

A potential short-coming of the approach is the lack of haptic (texture, heft) and olfactory feedback (smells) that form an important part of the ontological qualities of material culture and where its absence may diminish the experience. In the specific case described here, that was of no concern as glazed pottery is odorless and does not offer differential textures. The weight/heft of the specimens was also irrelevant in our case.

The ease of use with personally available smartphone technology will allow a researcher to scan an object in a private collection or field setting, where removal is ethically, legally or physically impossible. Moreover, it allows the researcher to share the image with remotely located collaborating researchers who, after manipulation of the 3D model, can instruct the researcher in the field to make additional observations if required. This generates a higher level of collaborative potential than looking at objects via Facetime, Zoom or similar technologies.

While our primary motivation related to research collaboration, these 3D models can also be generated 'on the fly' to be used in online tutorials to demonstrate certain aspects, allowing students to independently look at the examples as well.

## 5. Conclusions

The low-cost example described here can be used anywhere as long as a network connection can be accessed via WiFi or a mobile/cell phone network to upload the images. All subsequent processes can occur asynchronously. If the need arises, a series of images can be taken and uploaded and processed at a later point in time. Once uploaded and (automatically) processed, the images are located on the server and can be readily accessed and interactively manipulated anywhere in the world via a shared URL. The low-cost nature with widely owned technology (i.e. a camera-equipped smartphone) and internet access, as well as a mobility independent of fixed laboratory set-ups, makes the technology eminently suitable for the collection and dissemination of workable 3D models of items of material culture as encountered in small museums, private collections, or the field, where removal is ethically, legally or physically impossible.

**Appendix A: Server=-based 3D models generated for the project**

Three-dimensional models were generated for the following ten plates. The URLs point to the models residing on the Kiri Engine server. All links accessed on 8 December 2024.

| ID | Plate | URL |
|---|---|---|
| A1-1 | 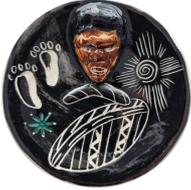 | Whole plate: https://www.kiriengine.app/share/ShareModel?code=W2ZK8I&serialize=31ecdf591301459bb4b072214db7ce85<br><br>Head only: https://www.kiriengine.app/share/ShareModel?code=W2ZK8I&serialize=275f4f9fd5f246738a4efa4421a2b037 |
| A1-2 | 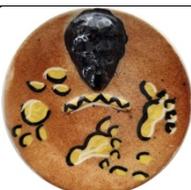 | Whole plate: https://www.kiriengine.app/share/ShareModel?code=W2ZK8I&serialize=0791880070c1425ca45d42c52f973bc9<br><br>Head only: https://www.kiriengine.app/share/ShareModel?code=W2ZK8I&serialize=4b260ae4e2394b85b3b6ec662ae53286 |
| A2-6 | 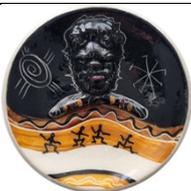 | Whole plate: https://www.kiriengine.app/share/ShareModel?code=W2ZK8I&serialize=927df77134674f0a992cf1b54c0edf75<br><br>Head only: https://www.kiriengine.app/share/ShareModel?code=W2ZK8I&serialize=e84e1ff34b8446229a3af99b7b189475 |
| A2-7 | 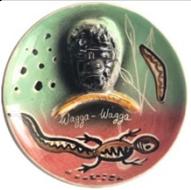 | Whole plate: https://www.kiriengine.app/share/ShareModel?code=W2ZK8I&serialize=5773c8ebb9e64fa5b06b38834eda00fb<br><br>Head only: https://www.kiriengine.app/share/ShareModel?code=W2ZK8I&serialize=cf2ec43fedec4cd6a1b769062071085b |
| A2–13 | 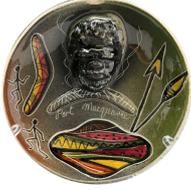 | Whole plate: https://www.kiriengine.app/share/ShareModel?code=W2ZK8I&serialize=9355474fa5a749918df19e4c18f555a3<br>Head only: https://www.kiriengine.app/share/ShareModel?code=W2ZK8I&serialize=632d60500ab04e489102471740225a67 |
| A2-16 | 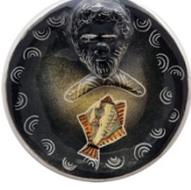 | Whole plate: https://www.kiriengine.app/share/ShareModel?code=W2ZK8I&serialize=6ab90e2cdde6421b932782ff7b894d80<br><br>Head only: https://www.kiriengine.app/share/ShareModel?code=W2ZK8I&serialize=977e0289fdb74b829cdaa1ad9e0ad9c6 |
| B1 | 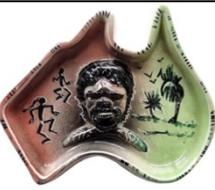 | Whole plate: https://www.kiriengine.app/share/ShareModel?code=W2ZK8I&serialize=ce761ec4f27e4db3813535c93d322b11<br><br>Head only: https://www.kiriengine.app/share/ShareModel?code=W2ZK8I&serialize=958ce9ed33aa45d3afdda41a92de007d |

| ID | Plate | URL |
|---|---|---|
| B6 | 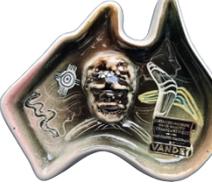 | Whole plate: https://www.kiriengine.app/share/ShareModel?code=W2ZK8I&serialize=6582b4a7aa5a4fe5b967e7a65fb25661<br>Head only: https://www.kiriengine.app/share/ShareModel?code=W2ZK8I&serialize=632d60500ab04e489102471740225a67 |
| D1 | 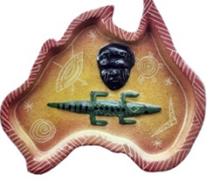 | Whole plate: https://www.kiriengine.app/share/ShareModel?code=W2ZK8I&serialize=903f0caaac9248139000e3a2dbd85c84<br>Head only: https://www.kiriengine.app/share/ShareModel?code=W2ZK8I&serialize=f5c9d87dabe643deb57514e0031db523 |
| D2 | 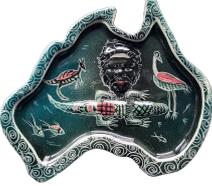 | Whole plate: https://www.kiriengine.app/share/ShareModel?code=W2ZK8I&serialize=c20c35882d434d869ab2f41d6b834433<br><br>Head only: https://www.kiriengine.app/share/ShareModel?code=W2ZK8I&serialize=a0aa6313633d42fa9e878517a7c03e04 |

**Appendix B: Time taken for various model development steps (hh:mm:ss)**

|  | A1-2 | A2-6 | D1 | B6 | A2-13 |
|---|---|---|---|---|---|
| Image acquisition |  |  |  |  |  |
| Photography | 00:03:09 | 00:02:45 | 00:03:38 | 00:02:59 | 00:03:23 |
|  |  |  |  |  |  |
| Processing |  |  |  |  |  |
| Upload | 00:19:52 *) | 00:15:26 **) | 00:08:18 | 00:09:54 | 00:07:52 |
| Queueing | 00:00:00 | 00:02:05 | 00:03:23 | 00:00:56 | 00:00:14 |
| Remote Model Processing | 00:49:00 | 00:11:52 | 00:14:06 | 00:12:54 | 00:44:20 |
| Total time to useable model | 01:12:01 | 00:32:08 | 00:29:25 | 00:26:43 | 00:55:49 |
|  |  |  |  |  |  |
| Crop to plate margin |  |  |  |  |  |
| manual adjustment side 1 | 00:00:23 | 00:00:27 | 00:00:20 | 00:00:23 | 00:00:21 |
| Processing side 1 | 00:01:45 | 00:01:05 | 00:06:38 | 00:09:13 | 00:01:09 |
| manual adjustment side 2 | 00:00:32 | 00:00:21 | 00:00:23 | 00:00:25 | 00:00:26 |
| Processing side 2 | 00:01:39 | 00:06:52 | 00:10:23 | 00:06:58 | 00:01:19 |
| manual adjustment side 3 | 00:00:26 | 00:00:20 | — | 00:00:22 | 00:00:19 |
| Processing side 3 | 00:07:05 | 00:09:09 | — | 00:06:52 | 00:00:45 |
| manual adjustment side 4 | 00:00:34 | 00:00:21 | — | 00:00:31 | 00:00:25 |
| Processing side 4 | 00:05:40 | 00:06:57 | — | 00:05:54 | 00:00:37 |
|  |  |  |  |  |  |
| Crop to head margin |  |  |  |  |  |
| manual adjustment side 1 | 00:00:20 | 00:00:22 | 00:00:18 | 00:00:15 | 00:00:19 |
| Processing side 1 | 00:08:21 | 00:05:27 | 00:02:02 | 00:05:49 | 00:01:05 |
| manual adjustment side 2 | 00:00:22 | 00:00:27 | 00:00:24 | 00:00:27 | 00:00:16 |
| Processing side 2 | 00:00:41 | 00:06:27 | 00:09:04 | 00:08:38 | 00:00:45 |
| manual adjustment side 3 | 00:00:21 | 00:00:30 | 00:00:24 | 00:00:27 | 00:00:20 |
| Processing side 3 | 00:00:53 | 00:06:44 | 00:00:58 | 00:08:55 | 00:00:45 |
| manual adjustment side 4 | 00:00:24 | 00:00:26 | 00:00:21 | 00:00:31 | 00:00:19 |
| Processing side 4 | 00:05:59 | 00:08:39 | 00:00:44 | 00:06;40 | 00:00:45 |
| file renaming and deletion of superfluous files | 00:00:12 | 00:00:10 | 00:00:12 | 00:00:14 | 00:00:15 |
| Total time | 01:47:38 | 01:26:52 | 01:01:36 | 01:22:37 | 01:05:59 |

*) Wifi to Broadband, 12.77 mbps upload; **) phone upload to 5G cell phone network;

**Author Contributions**: Conceptualization DHRS; Methodology DHRS; Data Curation DHRS; Formal Analysis DHRS; Writing – Original Draft Preparation DHRS; Writing – Review & Editing DHRS & SH; Visualization DHRS.

**Data availability**: The data can be accessed from authors' institutional research depository via this URL: https://doi.org/**10.26189/d0d4ab7a-4003-4d45-92fb-6976c2f5096d**.